\documentclass[10pt]{article}
\usepackage{amssymb} 

\def \ccomma{\raise 2pt\hbox{,}} 
\def \D {\hbox{d}}

\def \matU {\mathbb{U}}
\def \matV {\mathbb{V}}

\def \PVI    {{\rm P_{\rm VI}}}

\def \barQ {\overline{Q}}
\def \bfF {{\bf F}}
\def \bfN {{\bf N}}

\def \barz {{\bar z}}
\def \ch {c_{\rm h}}
\def \cz {c_{\rm z}}
\def \cq {c_{\rm q}}

\def \Hu{\textit{\rm H}}  
\def \HVI {\Hu_{\rm VI}}
\def \diag {\mathop{\rm diag}\nolimits}

\def\auu{a}
\def \A     {e} 
\def \Mzero {e_0}
\def \Mone  {e_1}

\title{Surfaces de Bonnet et \'equations de Painlev\'e}

\author{\textsc{Robert Conte${}^{1,2}$ 
} 
\\ \noindent 1.
    Centre de math\'ematiques et de leurs applications,  
\\ \'Ecole normale sup\'erieure de Cachan, CNRS, Universit\'e Paris-Saclay,
\\ 61, avenue du Pr\'esident Wilson, F--94235 Cachan Cedex, France.
\smallskip
\\ \noindent 2.
Department of mathematics,
The University of Hong Kong,
\\ Pokfulam road, Hong Kong.
\smallskip
\\ \noindent Courriel~: Robert.Conte@cea.fr
}

\begin{document}

\maketitle

\medskip
\begin{center}
{\small Soumis le 18 avril 2016; r\'evis\'e le 18 octobre 2016;
accept\'e le 19 octobre 2016\\ 
}
\end{center}

\noindent{\bf R\'esum\'e} \vskip 0.5\baselineskip \noindent

Nous montrons que les \'equations du rep\`ere mobile des surfaces de Bonnet
conduisent \`a 
une paire de Lax matricielle isomonodromique
d'ordre deux pour la sixi\`eme \'equation de Painlev\'e.

\medskip

\noindent{\bf Abstract} \vskip 0.5\baselineskip \noindent
{\bf Title. Bonnet surfaces and Painlev\'e equations}

We show that the moving frame equations of Bonnet surfaces
can be extrapolated to a second order, isomonodromic matrix Lax pair of the 
sixth Painlev\'e equation.

\vskip 0.5\baselineskip


\section*{Abridged English version}

In 1867, the geometer Pierre-Ossian Bonnet stated and solved the following problem: 
to determine all surfaces in $\mathbb{R}^3$ 
that can be mapped onto a given surface 
while conserving the principal radii of curvature.
Among the numerous solutions,
one important class, known as \textit{Bonnet surfaces},
depends on six arbitrary constants and is characterized by
the differential equation (\ref{eqBonnetsurfaces})
\cite[\S 11 p.~84 Eq.~(52)]{Bonnet1867}
for the mean curvature $H$.
Since this equation is solved \cite{BEK1997} by the Hamiltonian of a
particular sixth equation $\PVI$ of Painlev\'e,
the moving frame equations define a linear representation of this Hamiltonian.
In this Note,
we remove the restriction (\ref{eqBonnetContraintes-thetaj}),
express the moving frame equations with $\PVI$ instead of its Hamiltonian,
and finally normalize the result according to the prescription of Schlesinger.
Our final result is an isomonodromic matrix Lax pair of $\PVI$,
either very symmetric but restricted to $\theta_\infty\not=0$,
see (\ref{eqLax-PVI-codim0-balanced-meromorphic}),
or polynomial in the four $\theta_j$'s but less nice-looking,
see (\ref{eqLax-PVI-codim0-unbalanced-holomorphic}).

\section{Rappels sur les surfaces de Bonnet}

Parmi les surfaces de $\mathbb{R}^3$ applicables sur une surface donn\'ee
avec conservation des courbures principales,
il existe une classe remarquable, 
d\'ependant de six constantes arbitraires
et connue sous le nom de \textit{surfaces de Bonnet},
qui est caract\'eris\'ee par l'\'equation diff\'erentielle 
\cite[\S 11 p.~84 Eq.~(52)]{Bonnet1867}
\begin{eqnarray}
& &
\frac{1}{2} (\log H')''
-\left(\frac{\alpha}{\sin \alpha \xi}\right)^2 \frac{H' + H^2}{H'} + H'=0,
\label{eqBonnetsurfaces}
\end{eqnarray} 
o\`u 
$\xi$  d\'esigne une certaine fonction des coordonn\'ees conformes $z$, $\barz$,
$H(\xi)$ la courbure moyenne,
et $\alpha$ une constante \'eventuellement nulle.

Ce n'est que cent trente ans plus tard \cite{BEK1997} que pour $\alpha$ non nul 
sa solution $H$ sera
reconnue \'egale \`a l'hamiltonien $\HVI$ 
\cite[Eq.~(3)]{PaiCRAS1906} \cite[$t$ page 341]{ChazyThese}
\cite{MalmquistP6},
\begin{eqnarray}
& & 
\Hu_{\rm VI}
=\frac{x(x-1) {u'}^2}{4 u(u-1)(u-x)} 
\\ & & 
+\frac{1}{4 x(x-1)}
\left[
  \theta_\infty^2 \left(\frac{1}{2}-u\right) 
+ \theta_0^2      \left(\frac{1}{2}-\frac{x}{u}\right)
\right. \nonumber\\ & & \phantom{1234567890}\left.
+ \theta_1^2      \left(\frac{1}{2}-\frac{x-1}{u-1}\right)
+(\theta_x-1)^2   \left(\frac{1}{2}-\frac{x(x-1)}{u-x} -x \right)
\right],
\nonumber
\end{eqnarray}
de la sixi\`eme fonction de Painlev\'e $\PVI$, d\'efinie par
\begin{eqnarray}
& & {\hskip -20.0 truemm}
\frac{\D^2 u}{\D x^2}=
 \frac{1}{2} \left[\frac{1}{u} + \frac{1}{u-1} + \frac{1}{u-x} \right] u'^2
- \left[\frac{1}{x} + \frac{1}{x-1} + \frac{1}{u-x} \right] u'
\nonumber \\ & & \phantom{12345} {\hskip -20.0 truemm}
+ \frac{u (u-1) (u-x)}{2 x^2 (x-1)^2}
  \left[\theta_\infty^2 - \theta_0^2 \frac{x}{u^2} + \theta_1^2 \frac{x-1}{(u-1)^2}
        + (1-\theta_x^2) \frac{x (x-1)}{(u-x)^2} \right],
\end{eqnarray} 
avec toutefois trois 
contraintes entre les quatre param\`etres $\theta_j$,
\begin{eqnarray}
& &
\theta_\infty=0,\ 
\theta_0^2=
\theta_1^2,\ 
\theta_x^2=1.
\label{eqBonnetContraintes-thetaj}
\end{eqnarray} 

La consid\'eration des vari\'et\'es riemaniennes $\mathbb{R}^3(c)$,
que nous adoptons d\'esor\-mais en lieu et place de $\mathbb{R}^3$, 
permet 
d'introduire un param\`etre
suppl\'ementaire $c$ \cite{Springborn} 
dans les \'equations du rep\`ere mobile
et donc 
de rel\^acher une contrainte 
puisque $\theta_0^2-\theta_1^2=4 c$, 
\begin{eqnarray}
& &
\D \sigma=(\matU \D z + \matV \D \barz) \sigma,\ 
\\ & & {\hskip -22.0truemm}
\matU=\pmatrix{
 (1/4) U_z           & -Q e^{-U/2} \cr 
 (1/2) (H+c) e^{U/2} & -(1/4) U_z \cr },\
\matV=\pmatrix{
-(1/4) U_\barz & -(1/2) (H-c) e^{U/2} \cr 
\barQ e^{-U/2} & (1/4) U_\barz \cr},
\end{eqnarray}
o\`u $e^U$, $H$, $Q$, $\barQ$ sont les coefficients 
des deux formes quadratiques fondamentales
\begin{eqnarray}
& &
{\rm I}=<\D \bfF,\D \bfF>=e^U \D z \ \D \barz, 
\\
& &
{\rm II}=-<\D\bfF,\D\bfN>=Q \D z^2 + e^U H \D z \ \D\barz + \barQ \D\barz^2.
\end{eqnarray} 

\section{Repr\'esentation lin\'eaire de $\HVI$ et de $\PVI$}

Apr\`es une transformation conforme, 
les valeurs de Bonnet \cite[\S 11]{Bonnet1867}
\begin{eqnarray}
& & {\hskip -15.0 truemm}
\left\lbrace 
\begin{array}{ll}
\displaystyle{
x=\frac{1}{1-e^{4 \cz (z+\barz)}},\ 
}\\ \displaystyle{
e^U =- \frac{32 \cz \cq}{\ch} \frac{\D x} {\D H},\
H = \frac{2 \ch \cz}{\cq}Y,\ Y= x(x-1) \Hu_{\rm VI},\ 
}\\ \displaystyle{
    Q=\frac{4 \cq}{\ch} 
\frac{2 \cz}{\sinh(2 \cz (z+\barz))}\frac{\sinh(2 \cz \barz)}{\sinh(2 \cz z)},\
\barQ=\frac{4 \cq}{\ch} 
\frac{2 \cz}{\sinh(2 \cz (z+\barz))}\frac{\sinh(2 \cz     z)}{\sinh(2 \cz \barz)},
}
\end{array}
\right.
\end{eqnarray}
o\`u les constantes complexes non nulles $\cq$, $\cz$, $\ch$ 
permettent de couvrir les conventions des divers auteurs \cite{CG2016}, 
d\'efinissent ainsi une repr\'esentation lin\'eaire de $\HVI$
par des matrices d'ordre deux de trace nulle,
\begin{eqnarray}
& & {\hskip -15.0 truemm}
\matU \D z + \matV \D \barz
=x (x-1) \frac{Y''}{Y'}(\cz \D \barz-\cz \D z)
\pmatrix{1 & 0 \cr 0 & -1 \cr}
\nonumber \\ & & {\hskip -15.0 truemm} \phantom{12345}
+ \sqrt{Y'} 
\pmatrix{0 & -S_1 \D     z -\frac{\displaystyle{Y -(\theta_0^2+\theta_1^2)/8}}{\displaystyle{Y'}} 4 \cz \D \barz \cr 
              S_2 \D \barz +\frac{\displaystyle{Y +(\theta_0^2+\theta_1^2)/8}}{\displaystyle{Y'}} 4 \cz \D z& 0 \cr},\	
\nonumber \\ & & {\hskip -15.0 truemm}
S_1 =\frac{2 \cz}{\sinh(2 \cz (z+\barz))}\frac{\sinh(2 \cz \barz)}{\sinh(2 \cz z)},\
S_2 =\frac{2 \cz}{\sinh(2 \cz (z+\barz))}\frac{\sinh(2 \cz     z)}{\sinh(2 \cz \barz)}\cdot
\end{eqnarray}

Il existe alors \cite{BEK1997} un changement de variables $(z,\barz) \to (x,t)$ 
\begin{eqnarray}
& &
x=\frac{1}{1-e^{4 \cz (z+\barz)}},\ 
t=\frac{1}{1-e^{4 \cz z}},\ 
\matU \D z + \matV \D \barz =L \D x + M \D t, 
\end{eqnarray} 
rendant les nouvelles matrices $L,M$ du rep\`ere mobile rationnelles en $x$ et $t$, 
\begin{eqnarray}
& & {\hskip -15.0 truemm}
L=                             -\frac{M_x}{t-x}+L_\infty,\
M=\frac{M_0}{t}+\frac{M_1}{t-1}+\frac{M_x}{t-x},\
M_\infty=-M_0-M_1-M_x.
\label{eq-Matrix-Lax-pair}
\end{eqnarray} 
Les quatre singularit\'es $t=\infty,0,1,x$ sont du type de Fuchs, 
$L_\infty$ et les r\'esidus $M_j$ 
 ne d\'ependent que de $x$, $Y$, $Y'$, $Y''$, $\theta_j$. 
C'est pr\'ecis\'ement \cite{SchlesingerP6} 
une paire de Lax matricielle isomonodromique de $\HVI$. 

La transformation birationnelle         
\cite[Eq.~(3)]{PaiCRAS1906} \cite[Table R]{Okamoto1980-II}
entre $\PVI$ et son Hamiltonien 
permet ensuite de la convertir en une paire de Lax de $\PVI$.
Apr\`es le changement de base d\'efini par la matrice de passage 
\begin{eqnarray}
& & 
P_1=\diag({Y'}^{1/4},{Y'}^{-1/4}),
\end{eqnarray}
les matrices $L$ et $M$ sont rationnelles en toutes les variables 
et polynomiales en $u'$, $\theta_0$, $\theta_1$.
Il est alors facile,
en exigeant seulement la conservation du degr\'e 
des $M_j$ et de $L_\infty$ en $u'$,
de lever les restrictions (\ref{eqBonnetContraintes-thetaj}),
lire les d\'etails dans un article en cours de r\'edaction.

Valable pour des $\theta_j$ quelconques, 
cette premi\`ere forme canonique de la paire de Lax, 
\begin{eqnarray}
& & {\hskip -35.0 truemm}
\left\lbrace
\begin{array}{ll} 
\displaystyle{
L=-\frac{u-x}{x(x-1)} M_\infty -\frac{M_x}{t-x}\ccomma\
M=\frac{M_0}{t}+\frac{M_1}{t-1}+\frac{M_x}{t-x}\ccomma\
}\\ \displaystyle{
M_\infty+M_0+M_1+M_x=0,
}\\ \displaystyle{
M_\infty=\frac{1}{4}\pmatrix{2 \auu & -4 \cr \auu^2-\theta_\infty^2 & -2 \auu\cr},
}\\ \displaystyle{
M_0=-\frac{1}{2(u-x)}\pmatrix{\Mzero & -2 u(u-x) \cr 
 \displaystyle{\frac{\Mzero^2-\theta_0^2(u-x)^2}{2 u (u-x)}} & -\Mzero \cr},\
}\\ \displaystyle{
}\\ \displaystyle{
M_1= \frac{1}{2(u-x)}\pmatrix{\Mone & -2 (u-1)(u-x) \cr 
 \displaystyle{\frac{\Mone^2-\theta_1^2(u-x)^2}{2 (u-1) (u-x)}} & -\Mone \cr},\
}\\ \displaystyle{
}\\ \displaystyle{
M_x     =\frac{1}{2}\pmatrix{-\Theta_x & 0 \cr 2 M_{x,21} & \Theta_x \cr},\
}\\ \displaystyle{
M_{x,21}=-\frac{
\A^2-(u-x)^2 ((\theta_\infty^2+\Theta_x^2-2 \auu\Theta_x) u(u-1)-\theta_0^2 (u-1)+\theta_1^2 u)}
{4 u(u-1)(u-x)^2},
}\\ \displaystyle{
\A=x(x-1)u'+\Theta_x u(u-1),\ \Theta_x^2=(\theta_x-1)^2,\ 
}\\ \displaystyle{
\Mzero=\A-(\Theta_x-\auu)u(u-x),
}\\ \displaystyle{
\Mone=\A-(\Theta_x-\auu)(u-1)(u-x),
}\\ \displaystyle{	
-4 \det M_j=\theta_j^2,\ j=\infty,0,1; -4 \det M_x=\Theta_x^2,														
}
\end{array}
\right.
\label{eqLax-PVI-codim0-unbalanced-rational}
\label{eqLax-PVI-codim0-unbalanced-holomorphic}
\end{eqnarray}
d\'epend d'une constante arbitraire $\auu$,
qu'il est loisible de choisir \'egale \`a $\Theta_x$ ou \`a $\pm\theta_\infty$.

Comme l'a montr\'e Schlesinger \cite[p.~105]{SchlesingerP6}, 
le r\'esidu $M_\infty$ est constant et, s'il est inversible, 
il existe un changement de base
permettant d'annuler le terme $L_\infty$
dans (\ref{eq-Matrix-Lax-pair})
et donc de d\'efinir la paire de Lax de mani\`ere unique.
D\'efini par la matrice de passage
$P_2 P_3$,
\begin{eqnarray}
& & {\hskip -15.0 truemm}
P_2=\pmatrix{2 & 2 \cr \auu-\theta_\infty & \auu+\theta_\infty \cr},
P_3=\pmatrix{g^{-1/2} & 0 \cr 0 & g^{1/2} \cr},\
\frac{g'}{g}=\theta_\infty\frac{u-x}{x(x-1)}\ccomma
\end{eqnarray}
il conduit pour $\theta_\infty$ non nul
\`a la deuxi\`eme forme canonique, 
particuli\`erement \'el\'egante,
\vfill\eject
\begin{eqnarray}
& & {\hskip -20.0 truemm}
\left\lbrace
\begin{array}{ll} 
\displaystyle{
\theta_\infty\not=0:\ 
L = -\frac{M_x}{t-x},\ 
M=\frac{M_0}{t}+\frac{M_1}{t-1}+\frac{M_x}{t-x},\
\Theta_x^2=(\theta_x-1)^2,\ 
}\\ \displaystyle{
M_\infty+M_0+M_1+M_x=0,
}\\ \displaystyle{
M_\infty =\frac{1}{2} \pmatrix{\theta_\infty & 0\cr 0 & -\theta_\infty \cr},\
}\\ \displaystyle{
M_{0,11}=\ \ \frac{u-1}{N}\left[\left(\A- \Theta_x                u(u-x)\right)^2-(u-x)^2(\theta_0^2+\theta_\infty^2 u^2)\right],\
}\\ \displaystyle{
M_{0,12}=\ \ \frac{u-1}{N}\left[\left(\A-(\Theta_x+\theta_\infty) u(u-x)\right)^2-(u-x)^2 \theta_0^2\right] g,\
}\\ \displaystyle{
M_{0,21}=-   \frac{u-1}{N}\left[\left(\A-(\Theta_x-\theta_\infty) u(u-x)\right)^2-(u-x)^2 \theta_0^2\right] g^{-1},\
}\\ \displaystyle{
M_{1,11}=-   \frac{u}  {N}\left[\left(\A- \Theta_x            (u-1)(u-x)\right)^2-(u-x)^2(\theta_1^2+\theta_\infty^2 (u-1)^2)\right],\
}\\ \displaystyle{
M_{1,12}=-   \frac{u}  {N}\left[\left(\A-(\Theta_x+\theta_\infty)(u-1)(u-x)\right)^2-(u-x)^2 \theta_1^2\right] g,\
}\\ \displaystyle{
M_{1,21}=\ \ \frac{u}  {N}\left[\left(\A-(\Theta_x-\theta_\infty)(u-1)(u-x)\right)^2-(u-x)^2 \theta_1^2\right] g^{-1},\
}\\ \displaystyle{
M_{x,11}=\ \ \frac{1}  {N}\left[      \A                                          ^2-(u-x)^2
  \left[(\theta_\infty^2+\Theta_x^2)u(u-1)-\theta_0^2 (u-1) + \theta_1^2 u)\right]\right],\
}\\ \displaystyle{
M_{x,12}=\ \ \frac{1}  {N}\left[      \A                                          ^2-(u-x)^2 
  \left[(\Theta_x+\theta_\infty)^2 )u(u-1)-\theta_0^2 (u-1) + \theta_1^2 u)\right]\right] g,\
}\\ \displaystyle{
M_{x,21}=-   \frac{1}  {N}\left[      \A                                          ^2-(u-x)^2 
  \left[(\Theta_x-\theta_\infty)^2 )u(u-1)-\theta_0^2 (u-1) + \theta_1^2 u)\right]\right] g^{-1},\
}\\ \displaystyle{
-4 \det M_j=\theta_j^2,\ j=\infty,0,1; -4 \det M_x=\Theta_x^2,
}
\end{array}
\right.
\label{eqLax-PVI-codim0-balanced-meromorphic}
\end{eqnarray}
avec les notations
\begin{eqnarray}
& & {\hskip -10.0 truemm}
\frac{g'}{g}= \theta_\infty\frac{u-x}{x(x-1)},\
\nonumber \\ & & {\hskip -10.0 truemm}
\A=x(x-1) u' + \Theta_x u(u-1),\
N=4 \theta_\infty u(u-1)(u-x)^2.
\end{eqnarray}
										
Ce r\'esultat,
qui cl\^ot nos pr\'ec\'edentes recherches \cite{LCM2003,C2006Kyoto}, 
est beaucoup plus simple et sym\'etrique que celui 
de Jimbo et Miwa 
\cite[Eq.~(C.47)]{JimboMiwaII} \cite{Cargese1996Mahoux}, 
lire une comparaison d\'etaill\'ee dans \cite{LCM2003} et dans
\cite[p.~211]{CMBook}.
L'avantage d\'ecisif que procure l'origine g\'eom\'etrique
de la repr\'esentation lin\'eaire
est l'inutilit\'e de devoir supposer une repr\'esentation des
quatre r\'esidus respectant la constance de leur d\'eterminant.

La structure de $\PVI$ ($u''$ polyn\^ome de degr\'e deux en $u'$)
pourrait laisser esp\'erer une paire de Lax encore plus simple
dont les r\'esidus seraient des polyn\^omes de degr\'e un en $u'$,
mais le calcul montre qu'il n'en est rien. 

Les quatre \'el\'ements non-diagonaux $M_{12}$, $M_{21}$ de
(\ref{eqLax-PVI-codim0-unbalanced-holomorphic})
et de
(\ref{eqLax-PVI-codim0-balanced-meromorphic})
ont chacun un seul z\'ero $t=f(u',u,x)$
(\`a condition de choisir $\auu=\pm \theta_\infty$ 
 dans (\ref{eqLax-PVI-codim0-unbalanced-holomorphic})),
ces quatre z\'eros sont chacun solution d'une \'equation $\PVI$
et ils sont donc reli\'es entre eux par des transformations birationnelles,
comme sch\'ematis\'e dans \cite[Eq.~(4.4)]{LCM2003}.
Le plus simple de ces \'el\'ements est
\begin{eqnarray}
& & 
(\ref{eqLax-PVI-codim0-unbalanced-holomorphic}):\
M_{12}=\frac{t-u}{t(t-1)}\cdot
\end{eqnarray}
L'\'elimination d'une des deux composantes de $\sigma$,
que ce soit dans (\ref{eqLax-PVI-codim0-unbalanced-holomorphic})
ou dans
(\ref{eqLax-PVI-codim0-balanced-meromorphic}),
engendre donc bien l'unique singularit\'e apparente 
($t=u$ dans l'exemple ci-dessus, $t=$ une autre fonction $\PVI$ dans les trois autres cas))
que poss\`ede la paire de Lax scalaire classique \cite{FuchsP6}. 

Enfin, la confluence classique \cite{PaiCRAS1906} de $\PVI$ 
vers les cinq autres \'equations de Painlev\'e
d\'efinit des paires de Lax tout aussi sym\'etriques,
dont certaines apparemment nouvelles.

\section*{Remerciements}

C'est un plaisir de remercier
l'Unit\'e mixte internationale UMI 3457 du Centre de recherches math\'ematiques de 
l'Universit\'e de Montr\'eal pour son soutien financier.


\vfill\eject
\end{document}